\documentstyle[preprint,aps]{revtex}
\begin{document}
\tightenlines
\draft
\preprint{
\parbox{4cm}{
\baselineskip=12pt
TMUP-HEL-9912\\
TIT/HEP-435\\ 
KEK-TH-659\\ 
\hspace*{1cm}
}}
\title{Dynamical Supersymmetry Breaking \\
with Gauged $U(1)_R$ Symmetry}
\author{ Noriaki Kitazawa $^a$ 
  \thanks{e-mail: kitazawa@phys.metro-u.ac.jp}, 
         Nobuhito Maru $^b$ 
  \thanks{e-mail: maru@th.phys.titech.ac.jp, JSPS Research Fellow}
     and Nobuchika Okada $^c$ 
  \thanks{e-mail: okadan@camry.kek.jp, JSPS Research Fellow}}  

\address{$^a$Department of Physics, Tokyo Metropolitan University,\\
         Hachioji, Tokyo 192-0397, Japan}
\address{$^b$Department of Physics, Tokyo Institute of Technology,\\
         Oh-Okayama, Meguro, Tokyo 152-8551, Japan}
\address{$^c$Theory Group, KEK, Tsukuba, Ibaraki 305-0801, Japan}
%
\maketitle
%
%
\vskip 2.5cm
\begin{center}
{\large Abstract}
\vskip 0.5cm
\begin{minipage}[t]{14cm}
\baselineskip=19pt
\hskip4mm
We propose a simple model of dynamical supersymmetry breaking 
 in the context of minimal supergravity with gauged $U(1)_R$ symmetry. 
The model is based on the gauge group $SU(2) \times U(1)_R$ 
 with three matters. 
Since the $U(1)_R$ symmetry is gauged, the Fayet-Iliopoulos D-term appears 
 due to the symmetry of supergravity. 
On the other hand, the superpotential generated dynamically 
 by the $SU(2)$ gauge dynamics leads to run away potential. 
Since the supersymmetric vacuum condition required by the D-term potential 
 contradicts the one required by the superpotential,
 supersymmetry is broken. 
The supersymmetry breaking scale is controlled by the dynamical scale of 
 the $SU(2)$ gauge interaction. 
We can choose the parameters in our model 
 for vanishing cosmological constant. 
Our model is phenomenologically viable with the gravitino mass 
 of order 1 TeV or 10 TeV. 
%
\end{minipage}
\end{center}
\newpage
The supersymmetric extension is one of the most promising way 
 to provide a solution to the gauge hierarchy problem 
 beyond the standard model \cite{review}. 
However, since none of the superpartners has been observed yet, 
 supersymmetry should be broken at low energies. 
The origin of the supersymmetry breaking still remains 
 as the one of the biggest mysteries in supersymmetric theories. 

The models of the spontaneous supersymmetry breaking at the tree level
 were proposed many years ago \cite{OR-F-I}. 
However, since these models had dimensionful parameters given by hand, 
 there was no explanation for the hierarchy between the scale 
 of the supersymmetry breaking and Planck scale. 
More complete model may be the model 
 in which the origin of the scale of the supersymmetry breaking 
 can be explained by the model itself. 
An example of such model is the dynamical supersymmetry breaking model 
\cite{dsb}. 
While this model have no dimensionful parameter from the beginning, 
 the dimensionful parameter is induced 
 by the non-perturbative gauge dynamics. 
It seems to be possible to extend such a model into the supergravity model, 
 if the four dimensional space-time is flat. 

In this paper,
 we propose a simple model of dynamical supersymmetry breaking
 in the context of the minimal supergravity with gauged $U(1)_R$ symmetry.
Our model is based on the gauge group $SU(2) \times U(1)_R$. 
Since the $U(1)_R$ symmetry is gauged, the Fayet-Iliopoulos D-term 
 appears due to the symmetry of supergravity. 
On the other hand,
 the non-perturbative effect of the $SU(2)$ gauge dynamics 
 generates the superpotential dynamically, 
 which leads to the run away potential. 
Since the supersymmetric vacuum condition required by the D-term potential 
 contradicts the one required by the superpotential, supersymmetry is
broken. 
The supersymmetry breaking scale is controlled 
 by the scale of the $SU(2)$ gauge dynamics. 
Analyzing the potential minimum,
 we find that the cosmological constant can vanish, 
 if the parameters in our model are appropriately chosen. 
The mass spectrum of the model is also discussed. 
The scalars with non-zero $U(1)_R$ charges
 get soft supersymmetry breaking masses  
 at the tree level by the vacuum expectation value of the D-term. 
These masses are the same order of the magnitude of the gravitino mass. 
On the other hand, for the gauginos 
 in the minimal supersymmetric standard model 
 we can consider two possibilities. 
One is to introduce the higher dimensional term
 in the gauge kinetic function. 
The other is to consider the anomaly mediation scenario \cite{anomaly} 
 without non-trivial gauge kinetic function. 
The gaugino masses are found to be the same order of the gravitino mass 
 or a few orders smaller than the gravitino mass 
 in the former case or the latter case, respectively. 

Our model is based on the gauge group $SU(2) \times U(1)_R$ 
 with the following matter contents.
\footnote{In the following, we do not discuss the cancellation 
 of the gauge anomaly $[U(1)_R]^3$ and the mixed gravitational anomaly 
 of $U(1)_R$. 
The discussion depends on the full particle contents of the theory, 
 and it is out of the main subject of this paper \cite{chamseddine}. 
Here, we simply assume that these anomalies are canceled, 
 if all particle contents are considered
 with the appropriate $U(1)_R$ charge assignment.} 
\begin{center}
\begin{tabular}{ccc}
 \hspace{1cm}& $~SU(2)$~ & $~U(1)_R~$  \\
$ Q_1 $   &  \bf{2}     & $ -1 $   \\
$ Q_2 $   &  \bf{2}     & $ -1 $   \\
$ S $     &  \bf{1}     & $ +4 $  
\end{tabular}
\end{center}
The general renormalizable superpotential at the tree level is   
\begin{eqnarray}
 W = \lambda S \left[ Q_1 Q_2 \right] \; ,  
\end{eqnarray} 
 where square brackets denote the contraction of the $SU(2)$ index 
 by the $\epsilon$-tensor, 
 $\lambda$ is a dimensionless coupling constant. 
We assume that $\lambda$ is real and positive in the following. 

It is known that the superpotential is generated dynamically  
 by non-perturbative (instanton) effect 
 of the $SU(2)$ gauge dynamics \cite{instanton}. 
The total effective superpotential is found to be 
\begin{eqnarray}
 W_{eff}= \lambda S \left[ Q_1 Q_2 \right] 
        + \frac{\Lambda^5}{\left[Q_1 Q_2 \right]} \; ,  
 \label{superpotential}
\end{eqnarray}
 where the second term is the dynamically generated superpotential, 
 and $ \Lambda $ is the dynamical scale of the $SU(2)$ gauge interaction. 
Note that the supersymmetric vacuum lies 
 at $ \langle S \rangle \rightarrow \infty$ and 
 $ \langle Q_1\rangle, \langle Q_2 \rangle \rightarrow 0$, 
 if only the F-term potential is considered. 
 
Next, let us consider the D-term potential. 
The gauged $U(1)_R$ symmetry is impossible
 in the globally supersymmetric theory, 
 since the generators of the $U(1)_R$ symmetry and supersymmetry
 do not commute with each other. 
On the other hand, in the supergravity theory the $U(1)_R$ symmetry 
 can be gauged as if it were a usual global symmetry 
 \cite{superconf,chamseddine}. 
However, there is a crucial difference that
 the Fayet-Iliopoulos D-term of the gauged $U(1)_R$ symmetry
 appears due to the symmetry of supergravity. 
This fact is easily understood
 by the standard formula for supergravity theories \cite{cremmer}. 
Using the generalized K\"ahler potential $G = K + \ln |W|^2$, 
 the D-term is given by $D = \sum_i q_i (\partial G/ \partial z_i) z_i $, 
 where $q_i$ is the $U(1)_R$ charge of the field $z_i$. 
Note that the contribution from the superpotential leads to 
 the constant term, since the superpotential has $U(1)_R$ charge 2. 
 
With the above particle contents, the D-term potential is found to be 
\begin{eqnarray}
 V_D = \frac{g_R^2}{2} \left( 
       4 S^\dagger S - Q_1^\dagger Q_1 - Q_2^\dagger Q_2 + 2 M_P 
       \right)^2  \; , 
\end{eqnarray}
 where $M_P = M_{pl}/\sqrt{8 \pi}$ is the reduced Planck mass, 
 $g_R$ is the $U(1)_R$ gauge coupling, 
 and the minimal K\"ahler potential, 
 $K= S^\dagger S + Q_1^\dagger Q_1 + Q_2^\dagger Q_2 $,  
 is assumed.
\footnote{
This assumption is justified by our final result with $\Lambda \ll M_P$  
 which means that the $SU(2)$ gauge interaction
 is weak at the Planck scale. 
}
Note that the supersymmetric vacuum condition required by the D-term
potential 
 contradicts the one required by the effective superpotential 
 of eq.(\ref{superpotential}). 
Therefore, supersymmetry is broken. 
This consequence remains correct, 
 if there is no other superfields which have negative $U(1)_R$ charges. 
We give some comments on this point in the final part of this paper. 

Let us analyze the total potential in our model. 
Here, note that the cosmological constant should vanish. 
This requirement comes
 not only from the observations of the present universe 
 but also from the consistency of our discussion. 
Since it is not clear whether the superpotential discussed above 
 can be dynamically generated even in the curved space, 
 the space-time should be flat for our discussion to be correct. 
Note that we cannot take the usual strategy,
 namely, adding a constant term to the superpotential, 
 since such a term is forbidden by the $U(1)_R$ gauge symmetry. 
It is a non-trivial problem whether we can obtain 
 the vanishing cosmological constant in our model. 

Assuming that the potential minimum lies on the D-flat direction 
 of the SU(2) gauge interaction, 
 we take the vacuum expectation values such that 
 $ \langle S \rangle = s $ and 
 $\langle Q_i^\alpha  \rangle = v \delta_i^\alpha$, 
 where $i$ and $\alpha$ denote the flavor and $SU(2)$ indices, respectively.
Here, we can always make $s$ and $v$ real and positive 
 by symmetry transformations. 
The total potential is given by 
\begin{eqnarray}
   V(v,s) &=& e^K 
   \left[ \left( \lambda v^2 + s W \right)^2 
   + 2 v^2 \left( \lambda s - \frac{\Lambda^5}{v^4} + W \right)^2  
   -3 W^2  \right]    
 \label{potential}       \\ \nonumber  
   &+& \frac{g_R^2}{2} \left( 4 s^2 - 2 v^2 +2 \right)^2  \; , 
\end{eqnarray}
 where $K$ and $W$ are the K\"ahler potential and superpotential, 
 respectively, which are given by 
\begin{eqnarray} 
 K &=& s^2 + 2 v^2  \; ,  \\ 
 W &=& \lambda s v^2 + \frac{\Lambda^5}{v^2} \; . 
\end{eqnarray}
Here, all dimensionful parameters are taken to be dimensionless 
 with the normalization $M_P=1$. 
The first line in eq.(\ref{potential}) comes from the F-term  
 (except for $W^2$ term) and the remainder is the D-term potential.  

Since the potential is very complicated, it is convenient 
 to make some assumptions for the values of parameters. 
First, assume that $g_R \gg \lambda, \Lambda$. 
Since the D-term potential is proportional to $g_R^2$ and positive definite,
 the potential minimum is expected for $V_D$ to be small as possible. 
If we assume $s \ll 1 $ and $v \sim 1$, the potential can be rewritten as 
\begin{eqnarray}
 V \sim  e^2 \left( \lambda^2  - 3 \Lambda^{10}\right)  \; . 
\end{eqnarray}  
It is found that $\lambda \sim \sqrt{3} \Lambda^5$ is required 
 in order to get the vanishing cosmological constant. 
 
Let us consider the stationary conditions of the potential. 
Using the assumptions $s \ll 1$ and $v = 1+y $ ($|y| \ll 1$), 
 the stationary conditions can be expanded with respect to $s$ and $y$.  
Considering the relations $ g_R \gg \lambda \sim \Lambda^5$, 
 the condition $\partial V/ \partial y =0$ leads to   
\begin{eqnarray}
  y \sim s^2 - \frac{e^2 \lambda^2}{2 g_R^2}  \; . 
 \label{stationary1}
\end{eqnarray}
Using this result, the expansion of the condition 
 $\partial V/ \partial s  =0$ leads to 
\begin{eqnarray}
 s \sim \frac{ \lambda \Lambda^5} {8 \lambda^2 - \Lambda^{10}} \; .
 \label{stationary2}
\end{eqnarray}
By the numerical analysis, 
 the above rough estimation is found to be a good approximation. 
The result of numerical calculations is the following. 
\begin{eqnarray}
 y &\sim &  4.7 \times 10^{-3} \; ,
\label{y-value}   \\ 
 s &\sim &  6.8 \times 10^{-2} \; .  
\label{s-value}
\end{eqnarray}
Here, we used the values of $\Lambda =10^{-3}$, $ \lambda \sim 1.8 \;
\Lambda^5$ 
 and $g_R = 10^{-12}$. 
For these values of the parameters, 
 we can obtain the vanishing cosmological constant. 
Note that the numerical values of eqs. (\ref{y-value}) and (\ref{s-value})
 are almost independent of the actual value of $ \Lambda $, 
 if the condition $g_R \gg \Lambda^5$ is satisfied and 
 the ratio $\lambda/ \Lambda^5$ is fixed. 
This can be seen in the approximate formulae 
 of eqs.(\ref{stationary1}) and (\ref{stationary2}). 
We can choose the value of $\Lambda $ 
 in order to get a phenomenologically acceptable mass spectrum. 

Now we discuss the mass spectrum in our model. 
Using the above values of parameters, 
 the gravitino mass is estimated as  
\begin{eqnarray} 
  m_{3/2} = \langle e^{K/2} \; W  \rangle 
  \sim  3.0 \times  \frac{\Lambda^5}{M_P^4}  \; . 
\end{eqnarray}
The gravitino mass contributes to the masses of scalar partners 
 via the tree level interactions of supergravity. 
Note that there is another contribution, 
 if scalar partners have non-zero $U(1)_R$ charges. 
In this case, they also get the mass 
 from the vacuum expectation value of the D-term, 
 and it is estimated as 
\begin{eqnarray}
  m_{Dterm}^2 = q \; g_R^2 \langle D \rangle
  \sim \left( 7.3 \times \frac{\Lambda^5}{M_P^4} \right)^2 q \; , 
\end{eqnarray}
 where $q$ is the $U(1)_R$ charge.
This mass squared is always positive
 for the scalar partners with positive $U(1)_R$ charges.
The mass is the same order of the magnitude of the gravitino mass. 
This is because $g_R$ is canceled out in the above estimation
 (see eq.(\ref{stationary1})). 
For gaugino masses, we can consider two cases. 
One is to introduce a gauge invariant higher dimensional term 
 $ S ([Q_1 Q_2])^2 / M_P^5 $ in the gauge kinetic function. 
In this case,
 gaugino masses are found to be the same order of the gravitino mass. 
The other is to consider the anomaly mediation 
 of supersymmetry breaking \cite{anomaly} 
 without the non-trivial gauge kinetic function.
\footnote{
 The higher dimensional term $ S ([Q_1 Q_2])^2 / M_P^5 $
 in the gauge kinetic function
 can be forbidden to all orders by the discrete symmetry.
} 
In this case, gaugino masses are given by the gravitino mass 
 times beta functions, which are a few orders smaller  
 than the gravitino mass. 
Considering the experimental bound on gaugino masses 
 in the minimal supersymmetric standard model \cite{PDG}, 
 the gravitino mass is taken to be of the order of 1 TeV or 10 TeV 
 in the former case or the latter case, respectively. 
From this phenomenological constraint, 
the dynamical scale of the $SU(2)$ gauge interaction is found to be 
 of the order of $10^{15}$ GeV for both cases. 
This means that we have to fine-tune $\lambda \sim 10^{-15}$ 
to have the vanishing cosmological constant at tree level. 
\footnote{This small Yukawa coupling is consistent with our discussion
 in the following sense. 
Since $S$ has the vacuum expectation value, the mass for $Q_i$ 
 is generated through the Yukawa coupling in eq.(\ref{superpotential}). 
The relation $\lambda \langle S \rangle \ll \Lambda $ is needed 
 not to change our result from the $SU(2)$ gauge dynamics.}
This fine-tuning is also necessary in order to get the soft supersymmetry 
breaking masses of the same order of the gravitino mass.

Finally, we give some comments. 
Our model has the same structure of the supersymmetry breaking model 
 with the anomalous $U(1)$ gauge symmetry \cite{anomalousU1}. 
In the model, the Fayet-Iliopoulos D-term is originated 
 from the anomaly of the  $U(1)$ gauge symmetry \cite{superstring}. 
On the other hand, in our model the origin of the D-term 
 is the symmetry of supergravity with the gauged $U(1)_R$ symmetry.  
The D-term appears even if the $U(1)_R$ gauge interaction is anomaly free. 

The mechanism of the supersymmetry breaking in our model can work 
 unless there are other superfields  with negative $U(1)_R$ charges. 
However, when our model is combined with the visible sector, for example, 
 the minimal supersymmetric standard model,  
 it is highly non-trivial whether all the gauge anomalies  
 can be canceled out with only semi-positive $U(1)_R$ charged superfields 
 in the visible sector \cite{chamseddine}. 
The easiest way to remain our discussion correct 
 is to give up the cancellations of all the anomalies,  
 and consider the Green-Schwarz mechanism \cite{green} 
 by introducing the dilaton field as it is done in the model 
 with anomalous $U(1)$ gauge symmetry \cite{anomalousU1}.  
In this case, one can construct a full model combined 
 with our hidden sector \cite{kitazawa}. 
Although new Fayet-Iliopoulos D-term appears due to the $U(1)_R$ gauge 
 anomaly, its magnitude is suppressed compared 
 with that of the gauged $U(1)_R$ symmetry. 
Hence, our results obtained above is little changed. 

Unfortunately, the introduction of the dilaton field in the model 
 causes new difficult problems such as the stabilization 
 of the dilaton potential, the vacuum expectation value 
 of the dilaton F-term and so on \cite{dilaton}. 
Therefore, it is likely expected to construct the model 
 with our supersymmetry breaking mechanism
 without the Green-Schwarz mechanism. 
Indeed, we can construct the anomaly free model with some extensions  
 of the model presented in this paper \cite{kitazawa}.  
\acknowledgments
We would like to thank Hitoshi Murayama for helpful comments. 
This work was supported in part by the Grant-in-aid for Science and Culture
Research form the Ministry of Education, 
Science and Culture of Japan (\#11740156, \#3400, \#2997). 
N.M and N.O. are supported by the Japan Society for the Promotion of Science 
for Young Scientists.
%

%
\end{document}